\documentclass{article}
\usepackage{spconf,graphicx}
\usepackage{amsmath,amssymb,amsfonts}

\usepackage[dvipsnames]{xcolor}
\usepackage{bm}			
\usepackage{cite} 
\usepackage[per-mode=symbol]{siunitx}    		
\DeclareSIUnit{\dBm}{dBm}	
\DeclareSIUnit{\sqrtW}{\ensuremath{\sqrt{\text{W}}}}
\usepackage{layouts}
\usepackage[compact]{titlesec}  
\usepackage[font=footnotesize]{caption}
\usepackage[font=footnotesize]{subcaption}      

\definecolor{mygray}{gray}{0.6}
\usepackage[    
                backgroundcolor=RDlightgreen!70,
                textcolor=black,
                bordercolor=RDgreen,
                textsize=tiny]{todonotes}
\makeatletter
\renewcommand{\todo}[2][]{\tikzexternaldisable\@todo[#1]{#2}\tikzexternalenable}
\makeatother

\usepackage{pgfplots}
\usepackage{ifthen}
\newcommand{\externalizeFigures}{false}
\ifthenelse{\equal{\externalizeFigures}{true}}
{
  \pgfplotsset{compat=newest}
  \usepgfplotslibrary{external}
  \tikzexternalize[prefix=externalized/]

  \usepackage{shellesc}
}
{}
  \pgfplotsset{compat=newest}

\usetikzlibrary{positioning,spy}                
\usetikzlibrary{datavisualization.polar}        
\usepgfplotslibrary{polar}

\usepackage{pgfplots}
\pgfplotsset{compat=newest}
\usetikzlibrary{plotmarks}
\usetikzlibrary{arrows.meta}

\usepgfplotslibrary{patchplots}
\usepackage{grffile}
\usepackage{amsmath}
\usetikzlibrary{calc}
\usepackage{tikzscale}		

\usepackage[xindy]{glossaries}
\makenoidxglossaries%
\newacronym{2d}{2D}{two-dimensional}
\newacronym{3d}{3D}{three-dimensional}
\newacronym{aoa}{AOA}{angle-of-arrival}
\newacronym{aod}{AOD}{angle-of-departure}
\newacronym{asic}{ASIC}{application specific integrated circuit}
\newacronym{awgn}{AWGN}{additive white Gaussian noise}
\newacronym{bf}{BF}{beamformer}
\newacronym{ce}{CE}{channel estimation}
\newacronym{ced}{CED}{cumulative energy density}
\newacronym{cdf}{CDF}{cumulative distribution function}
\newacronym{csi}{CSI}{channel state information}
\newacronym{crlb}{CRLB}{Cram\'er-Rao lower bound}
\newacronym{dc}{DC}{direct current}
\newacronym{dm}{DM}{diffuse multipath}
\newacronym{dmc}{DMC}{diffuse multipath component}
\newacronym{eh}{EH}{energy harvesting}
\newacronym{eirp}{EIRP}{equivalent isotropically radiated power}
\newacronym{em}{EM}{electromagnetic}
\newacronym{en}{EN}{energy neutral}
\newacronym{e2e}{E2E}{End-to-End}
\newacronym{epu}{EPU}{edge processing unit}
\newacronym{fa}{FA}{federation anchor}
\newacronym{jcrs}{JCRS}{joint communication and radar sensing }
\newacronym{ic}{IC}{integrated circuit}
\newacronym{iot}{IoT}{Internet of things}
\newacronym{id}{ID}{information decoding}
\newacronym{isac}{ISAC}{integrated sensing and communication}
\newacronym{ism}{ISM}{industrial, scientific and medical}
\newacronym{kpi}{KPI}{key performance indicator}
\newacronym{lca}{LCA}{life cycle assessment}
\newacronym{lis}{LIS}{large intelligent surface}
\newacronym{liion}{Li-ion}{lithium-ion}
\newacronym{los}{LoS}{line-of-sight}
\newacronym{lmo}{LMO}{lithium ion manganese oxide}
\newacronym{mc}{MC}{Monte Carlo}
\newacronym{mf}{MF}{matched filter}
\newacronym{mmse}{MMSE}{minimum mean square error}
\newacronym{mmwave}{mmWave}{millimeter wave}
\newacronym{mimo}{MIMO}{multiple-input multiple-output}
\newacronym{miso}{MISO}{multiple-input single-output}
\newacronym{nlos}{NLoS}{non-line-of-sight}
\newacronym{xlmimo}{XL-MIMO}{extremely large-scale \acrshort{mimo}}
\newacronym{mrc}{MRC}{maximum ratio combining}
\newacronym{mrt}{MRT}{maximum ratio transmission}
\newacronym{ofdm}{OFDM}{orthogonal frequency-division multiplexing}
\newacronym{ota}{OTA}{over-the-air}
\newacronym{nb}{NB}{narrowband}
\newacronym{nr}{NR}{New Radio}
\newacronym{pa}{PA}{power amplifier}
\newacronym{pdp}{PDP}{power delay profile}
\newacronym{per}{PER}{packet error rate}
\newacronym{pl}{PL}{path loss}
\newacronym{pg}{PG}{path gain}
\newacronym{pc}{PC}{pilot count}
\newacronym{pw}{PW}{planar wavefront}
\newacronym{rcs}{RCS}{radar cross-section}
\newacronym{re}{RE}{radio element}
\newacronym{rf}{RF}{radio frequency}
\newacronym{rfid}{RFID}{\acrshort{rf} identification}
\newacronym{ris}{RIS}{reflective intelligent surface}
\newacronym{rms}{RMS}{root mean square}
\newacronym{rss}{RSS}{received signal strength}
\newacronym{rssi}{RSSI}{received signal strength indicator}
\newacronym{rw}{RW}{RadioWeave}
\newacronym{sa}{SA}{synthetic aperture}
\newacronym{sdg}{SDG}{Sustainable Development Goal}
\newacronym{sdn}{SDN}{software-defined network}
\newacronym{sinr}{SINR}{signal-to-interference-plus-noise ratio}
\newacronym{siso}{SISO}{single-input single-output}
\newacronym{slam}{SLAM}{simultaneous localization and mapping}
\newacronym{smc}{SMC}{specular multipath component}
\newacronym{snr}{SNR}{signal-to-noise ratio}
\newacronym{s-parameter}{S-parameter}{scattering parameter}
\newacronym{svd}{SVD}{singular value decomposition}
\newacronym{sw}{SW}{spherical wavefront}
\newacronym{swipt}{SWIPT}{simultaneous wireless information and power transfer}
\newacronym{tdd}{TDD}{time division duplexing}
\newacronym{tdoa}{TDOA}{time-difference-of-arrival}
\newacronym{toa}{TOA}{time-of-arrival}
\newacronym{tosm}{TOSM}{through–open–short–match}
\newacronym{ue}{UE}{user equipment}
\newacronym{uhf}{UHF}{ultra high frequency}
\newacronym{ula}{ULA}{uniform linear array}
\newacronym{upa}{UPA}{uniform planar array}
\newacronym{ura}{URA}{uniform rectangular array}
\newacronym{uwb}{UWB}{ultrawideband}
\newacronym{vna}{VNA}{vector network analyzer}
\newacronym{wb}{WB}{wideband}
\newacronym{wpt}{WPT}{wireless power transfer}
\newacronym{wsn}{WSN}{wireless sensor network}
\newacronym{xets}{XETS}{cross exponentially tapered slot}
\newacronym{zf}{ZF}{zero forcing}
\newacronym{csp}{CSP}{contact service point}
\newacronym{ecsp}{ECSP}{edge computing service point}
\newacronym{ap}{AP}{access point}

\usepackage{balance}
\usepackage{flushend}
\usepackage{multirow}

\definecolor{RDlightgreen}{RGB}{141 192 69}
\definecolor{RDgreen}{rgb}{0.3647, 0.4275, 0.2667}
\definecolor{RDdarkgreen}{rgb}{0.2196, 0.2196, 0.2196}
\definecolor{RDmaroon}{rgb}{.522,.22,.353} %

\newcommand{\mrm}[1]{ \mathrm{#1} }

\newcommand{ \trp}{\mathsf{T}}

\newcommand{\complexset}[2]{ \mathbb{C}^{#1 \times #2}  }

\newlength\figureheight
\newlength\figurewidth 
\newlength\hspaceing
\newlength\plotHeight
\newlength\plotWidth

\def\BibTeX{{\rm B\kern-.05em{\sc i\kern-.025em b}\kern-.08em
    T\kern-.1667em\lower.7ex\hbox{E}\kern-.125emX}}

\title{Bistatic MIMO Radar Sensing of Specularly Reflecting Surfaces for Wireless Power Transfer}
\name{
Benjamin J.\,B. Deutschmann,   
Maximilian Graber,   
Thomas Wilding,   
Klaus Witrisal     
\thanks{The project has received funding from the European Union’s Horizon 2020 research and innovation program under grant agreement No 101013425.}
}

\address{Graz University of Technology, Austria 
}
\begin{document}
\ninept
\setlength{\abovedisplayskip}{6pt}
\setlength{\belowdisplayskip}{6pt}
\maketitle
\begin{abstract}
Geometric environment information aids future distributed radio infrastructures in providing services, such as ultra-reliable communication, positioning, and \gls{wpt}.
An a priori known environment model cannot always be assumed in practice.
This paper investigates the capabilities of detecting specularly reflecting surfaces in a bistatic \gls{mimo} radar setup operating at sub-10\,GHz frequencies.
While rough surfaces generate diffuse reflections originating from their actual position, flat surfaces act like ``mirrors,'' causing directive reflections that virtually originate ``behind'' them.
Despite these propagation characteristics, we can estimate the locations of flat metal walls from reflections originating at their surface using \gls{sa} measurements.
The performance gain achievable by exploiting this 
environment information is analyzed by evaluating \gls{wpt} capabilities in a geometry-based beamforming setup.
We show that it is possible to predict \gls{csi} with a geometric channel model. 
Our geometry-based beamformer suffers an efficiency loss of only 1.1\,dB compared with a reciprocity-based beamformer given perfect \gls{csi}.
\end{abstract}
%

\glsresetall

\section{Introduction}

Future distributed radio infrastructures like RadioWeaves~\cite{VanDerPerre2019} provide unprecedented potential for sensing in indoor environments even at sub-10\,GHz frequencies. 
Large numbers of distributed arrays operating in a cooperative fashion, e.g., as a multistatic \gls{mimo} radar 
\ifdefined\reduceSize  
    system, 
\else
     system~\cite{ThomaeWC2019}, 
\fi
may be capable of performing geometric environment mapping and object tracking, both of which aid services, such as ultra-reliable communication, positioning, and \gls{wpt}~\cite{D1_1}.
In this work, we explore sensing to aid a distributed radio infrastructure. 

The term \emph{remote sensing} is often defined as the acquisition of object-related data from a remote distance and is commonly used in fields surveying the surface of the Earth, e.g., geology or meteorology~\cite{Campbell11RemoteSensing}.
To meet the envisioned performance goals in 6G technologies, {\gls{isac}} 
with a single network architecture~\cite{Wei2022IntegratedSensing} as well as \gls{jcrs}~\cite{ThomaeEUCAP2021} are promising 
emerging fields.
Recent research on 6G systems operating in mmWave and THz bands promised outstanding indoor sensing capabilities given that building surfaces are rough w.r.t. the wavelength $\lambda$ and cause a mixture of \gls{dm} and \glspl{smc}~\cite{Rappaport2019-6G-100GHz}, with the surface roughness affecting the directivity of the scattered components \cite{KulmerPIMRC2018}. 
The low directivity of diffuse scattering has been exploited in~\cite{Aladsani2019mmWaveImaging} to infer positions of walls using \gls{sa} radar imaging in the upper mmWave frequency range of 
${220}$-$\SI{300}{\giga\hertz}$
with the ultimate goal of estimating the location of %
\ifdefined\reduceSize  
    a \gls{ue}.
\else
     a \gls{ue}, as well as in \cite{WenTWC2021} in the context of 5G systems.
\fi
\ifdefined\reduceSize  
\else
     Due to their high reflectivity, beamforming through metal walls is virtually impossible. 
     However, it has been demonstrated that diffraction at the edges of metal walls can be used to reach a \gls{ue} in a \gls{nlos} position~\cite{Zhang2018HardWallImaging}.
\fi
Antenna arrays operating at sub-10\,GHz frequencies are well suited for \gls{wpt} due to the large apertures involved and the low radiation levels achievable outside their focal region~\cite{D4_1}. 
In this frequency range, however, surfaces in buildings are typically large 
and flat w.r.t. the wavelength and thus cause specular reflections, commonly modeled by mirror sources~\cite{Leitinger2015}.
Specularly reflecting surfaces are usually difficult to detect from arbitrary directions due to their directional rather than isotropic reflections. Stealth technology exploits this concept, where geometric shapes often consist of flat facetted surfaces that intentionally forward-scatter incident radar signals away from the direction of anticipated radar locations~\cite{Willis2007,Zong2016}.
\Gls{wpt} can leverage directive specular reflections to increase the power budget by focusing multiple reflected beams at the position of a \gls{ue}~\cite{Deutschmann23ICC} or establish an \gls{smc} link for a \gls{ue} that is in \gls{nlos} conditions~\cite{Aladsani2019mmWaveImaging}. 

\begin{figure}
  \centering
  \setlength{\figurewidth}{0.95\columnwidth}
  \setlength{\figureheight}{1\columnwidth}
  \def\datapath{./figures/floorplan}
  \input{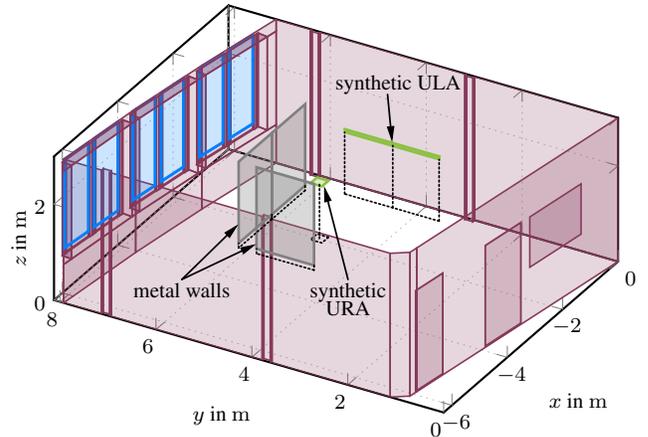}
  \vspace{-4mm}%
  \caption%
  {A \gls{3d} model of the measurement scenario:
  Bistatic \gls{sa} measurements are conducted using one $\frac{\lambda}{2}$-ULA mounted on a wall and a $\frac{\lambda}{4}$-URA in the horizontal plane at $f_\mrm{c}=\SI{3.79}{\giga\hertz}$.
  Metal walls have been placed in the environment to introduce specular reflections.}%
    \label{fig:scenario}%
    \vspace{-7mm}
\end{figure}


Waves reflected from point scatterers (possibly representing diffuse reflections) are modeled to originate from the location of a scatterer.
Specular reflections at flat surfaces, termed \glspl{smc}, are modeled to virtually originate from image sources obtained by mirroring the true source across the surface.
These surfaces act like mirrors and 
they are hard to detect from arbitrary directions. 
In this work, we aim to directly infer the location of a wall through a bistatic \gls{mimo} radar imaging scheme presented in~\cite{Spreng13MIMOradar} rather than estimating the position of a mirror source, which is conventionally done in \gls{slam}~\cite{Deissler10UWBslam} at sub-10\,GHz frequencies. 
We demonstrate that this environment information can subsequently be exploited in a geometry-based channel model to predict \gls{csi}, with the ultimate goal of performing efficient \gls{wpt}. 
In this context, the use of \gls{csi} for beamforming is commonly referred to as sensing-aided beam 
\ifdefined\reduceSize  
    prediction~\cite{Wei2022IntegratedSensing}.
\else
    prediction~\cite{Wei2022IntegratedSensing,Charan2022beamPrediction}.
\fi
The main contribution of this paper is to explore the potential of a distributed radio infrastructure inferring a geometric environment model and using it for efficient \gls{wpt} in a simultaneous multi-beam transmission.

The remainder of this paper is organized as follows. 
Section~\ref{sec:channel} introduces a channel model related to the \gls{sa} measurements described in Section~\ref{sec:meas-system}. 
Section~\ref{sec:sensing} describes an approach to infer wall locations in a bistatic \gls{mimo} system.
Section~\ref{sec:wpt} shows how this information can be used to perform efficient \gls{wpt}.
{The measurement data and code accompanying this paper are available at~\cite{GitLab} and~\cite{IEEE-dataport}.}

\section{Channel Model}\label{sec:channel}%
We use a geometry-based channel model for \gls{miso} systems to model the frequency domain channel vector $\bm{h}(\bm{p},f) \in \complexset{M}{1}$, for a frequency $f$ and a \gls{ue} position $\bm{p} = [p_x \, p_y \, p_z]^\trp$. 
The $m$\textsuperscript{th} element of the channel vector $[\bm{h}(\bm{p},f)]_m$ represents a forward transmission coefficient, i.e., a \gls{s-parameter} $S_{21}(f)$, from transmit antenna $m \in \{1 \, \hdots \, M\}$ to the \gls{ue} antenna. 
The channel vector is modeled as 
 the superposition of the channel vectors $\bm{h}_k(\bm{p},f)$ of $K$ \glspl{smc} 
\begin{align}\label{eq:channel-model-complete}
    \bm{h}(\bm{p},f) = \sum\limits_{k=1}^{K} \bm{h}_k(\bm{p},f) \, .
\end{align} 
Each \gls{smc} $k \in \{1 \, \dots \, K\}$ (including the \gls{los} with $k=1$) is modeled by means of a mirror source, obtained by mirroring all $M$ transmit antennas across the surface and computing the corresponding distances from the position of the $k$\textsuperscript{th} mirror source at position $\bm{p}_k$ to the \gls{ue} position $\bm{p}$ (see the appendix of~\cite{Deutschmann23ICC}). 
For simplicity, we only model first-order specular reflections. 
The elements of each \gls{smc} channel vector are accordingly modeled as~\cite{D4_1}
\begin{align}\label{eq:smc-channel-vector}
    \left[\bm{h}_k(\bm{p},f)\right]_{m} = \sqrt{G_{\mrm{t},m}} \sqrt{G_{\mrm{r}}} \frac{\lambda}{4\pi \lVert \bm{r}_{k,m} \rVert} e^{-j\frac{2\pi}{\lambda } \lVert \bm{r}_{k,m} \rVert}
\end{align}
which represents the Friis transmission equation formulated for power wave amplitudes.
$G_{\mrm{t},m}(\theta,\varphi)$ and $G_{\mrm{r}}(\theta,\varphi)$ are the gain patterns of the respective antennas in elevation and azimuth angles $(\theta,\varphi)$ 
in local spherical antenna coordinates, and $\bm{r}_{k,m} = \bm{p} - \bm{p}_{k,m}$ is the vector from transmit antenna $m$ of mirror source $k$ at $\bm{p}_{k,m}$ to the \gls{ue} position $\bm{p}$.

When transmitting with a total power $P_\mrm{t}$, the \gls{ue} receives a complex baseband amplitude, i.e., a phasor,
\begin{align}\label{eq:y_rx}
    \alpha(\bm{p},f,\bm{w}) = \bm{h}^\trp(\bm{p},f) \,  \bm{w} \sqrt{P_\mrm{t}} 
\end{align}
where $\bm{w} \in \complexset{M}{1}$ is a unit-vector of beamforming weights, i.e., $\lVert \bm{w} \rVert = 1$. 
The path gain defined as
\begin{align}
    PG(\bm{p},f,\bm{w}) = \frac{P_\mrm{r}}{P_\mrm{t}} = \frac{|\alpha(\bm{p},f,\bm{w})|^2}{P_\mrm{t}} 
\end{align}
is used to represent the power transmission efficiency as the ratio of received power $P_\mrm{r}$ to transmit power. We assume $P_\mrm{t}=\SI{1}{\watt}$ for the remainder of this paper.

\section{Measurement System}\label{sec:meas-system}%
We employ an \gls{sa} measurement testbed with two mechanical positioners to measure the channel vector elements $[\bm{h}(\bm{p}_n,f)]_m$ 
between antenna $m$ of a synthetic \gls{ula} and antenna $n \in \{1 \, \dots \, N\}$ of a synthetic \gls{ura}.
The \gls{ula} and \gls{ura} form a \gls{mimo} system, with the \gls{mimo} channel matrix $\bm{H}(f) = [\bm{h}(\bm{p}_1,f) \, \hdots \, \bm{h}(\bm{p}_N,f)] \in \complexset{M}{N}$ obtained by stacking the $N$ \gls{miso} channel vectors for each receive antenna $n$. 
The scenario is illustrated in Fig.\,\ref{fig:scenario}, with the \gls{ura} located between metal walls that generate strong \glspl{smc}. 
%
We use a Rohde \& Schwarz ZVA24 \gls{vna} in a two-port configuration to measure the transmission coefficient $S_{21}(f)$ between a transmit antenna $m$ connected to Port\,$1$ and a receiving antenna $n$ connected to Port\,$2$ (see~\cite{D1_2} for a description of the measurement system).
 We measure at $N_f=1000$ linearly spaced frequencies $f_i$ in a frequency band of ${3}$-$\SI{10}{\giga\hertz}$ .
The synthetic apertures are a $\frac{\lambda}{2}$-spaced $51$-\gls{ula} and a $\frac{\lambda}{4}$-spaced $(13 \times 13)$-\gls{ura} for a chosen carrier frequency of $f_\mrm{c} = \SI{3.79}{\giga\hertz}$, with $\lambda=\frac{c}{f_\mrm{c}}$.
%
\section{Radar Sensing of Surfaces}\label{sec:sensing}%
This section describes the implemented processing steps of radar imaging and surface estimation. 
We use the \gls{ura} as a transmitter and the \gls{ula} as a receiver to infer a geometric environment model.
\setlength{\abovedisplayskip}{0.1pt}
\setlength{\belowdisplayskip}{1pt}
\subsection{Radar Imaging}\label{sec:imaging}%
\begin{figure}
  \centering
  \includegraphics[width=\linewidth]{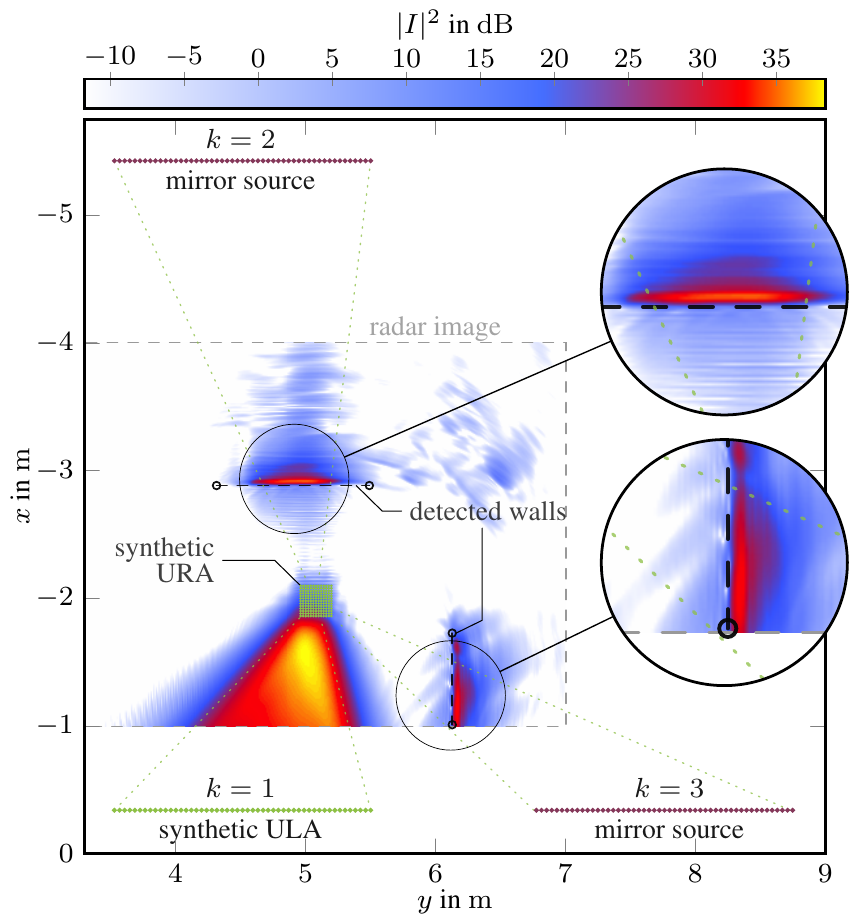}
  \vspace{-7mm}
  \caption{%
  The obtained bistatic \gls{mimo} radar image shows the received power of the imaging method (see Section~\ref{sec:imaging}) with the positions of walls estimated as described in Section~\ref{sec:hough}. 
  The positions of mirror sources $k\in\{2,3\}$ have been computed by mirroring the synthetic \gls{ula} ($k=1$) across the walls.}
    \label{fig:imaging}
    \vspace{-6mm}
\end{figure}

We employ the radar imaging scheme proposed in~\cite{Spreng13MIMOradar} to compute a reflectivity map \begin{align}\label{eq:imaging}
    I(\bm{p}) = \sum\limits_{i=1}^{N_f} \bm{w}_\mrm{r}^\trp(\bm{p},f_i) \, \bm{H}(f_i) \, \bm{w}_\mrm{t}(\bm{p},f_i) 
\end{align}
for candidate points $\bm{p}$ in a specified \gls{2d} spatial window of interest. 
The window is aligned with the 
vertical position of the \gls{ula}. 
\setlength{\abovedisplayskip}{6pt}
\setlength{\belowdisplayskip}{6pt}
The weight vectors for position-based beamforming, $\bm{w}_\mrm{r}(\bm{p},f_i)$ and $\bm{w}_\mrm{t}(\bm{p},f_i)$, for the receiving \gls{ula} and the transmitting \gls{ura}, respectively, are computed by applying \gls{mrt}, i.e.,
\begin{align}\label{eq:weights} 
    \bm{w}_\mrm{r}(\bm{p},f) = \frac{\bm{h}_\mrm{r}^*(\bm{p},f)}{\lVert \bm{h}_\mrm{r}(\bm{p},f) \rVert} \quad \text{and} \quad
    \bm{w}_\mrm{t}(\bm{p},f) = \frac{\bm{h}_\mrm{t}^*(\bm{p},f)}{\lVert \bm{h}_\mrm{t}(\bm{p},f) \rVert} \, ,
\end{align}
where the channel vectors $\bm{h}_\mrm{r}(\bm{p},f) \in \complexset{M}{1}$ and $\bm{h}_\mrm{t}(\bm{p},f)\in \complexset{N}{1}$ are given by~\eqref{eq:smc-channel-vector}, assuming isotropic gain patterns for simplicity and $K=1$, i.e., \gls{los}-only beamforming.
Through the channel definition in~\eqref{eq:smc-channel-vector}, spherical wavefront beamforming is inherently performed (cf.,~\cite{Vouras22IRSbeamforming,Aladsani2019mmWaveImaging,Deutschmann23ICC,Vouras23AdvancesSA}).
Fig.\,\ref{fig:imaging} shows the imaging results we obtain by evaluating \eqref{eq:imaging} for a window of positions $\bm{p}$ in the measurement scenario (see Fig.\,\ref{fig:scenario}). 
\ifdefined\reduceSize  
\else
    Comparing our results with the mmWave imaging performance in~\cite{Aladsani2019mmWaveImaging} (using a \gls{ue} equipped with a single antenna), we find that we have achieved a reasonable imaging result even when operating at sub-10\,GHz frequencies, at the cost of using a bistatic radar setup (i.e., arrays at both the transmitter and the receiver).
\fi
Note that $|I(\bm{p})|^2$ is a measure of the power received by the \gls{ula} from position $\bm{p}$ when simultaneously beamforming to the position $\bm{p}$ with the \gls{ura}, and coherently summing over the whole frequency band of ${3}$-$\SI{10}{\giga\hertz}$. 
The peak visible in the \gls{los} path between the two arrays, exhibiting \emph{smooth} edges, is due to direct ``illumination'' via the corresponding beams. 
\ifdefined\reduceSize  
\else
    The power decreases in the vicinity of the \gls{ura} because it is vertically located \SI{10}{\centi\meter} below the evaluated window.
\fi
The radar image shows \emph{sharp} edges at the locations of the metal walls (i.e., the specularly reflecting surfaces) 
and a gradually decaying power ``behind'' the walls.
\subsection{Edge Detection and Surface Estimation}\label{sec:hough}%
The sharp edges in the radar image are well-suited for an edge detection algorithm. 
We run the Canny edge detector~\cite[Sec.\,2.4]{Parker10ImageProcessing} 
on the radar image and subsequently employ the Hough transform~\cite[p.\,342\,ff.]{Parker10ImageProcessing} to transform the image into the Hough space. 
The peaks of the resulting Hough image are used to find the location, orientation, and extent of the detected edges.
Both are well-established methods in image processing. 
We use the \textsc{Matlab}\textsuperscript{\textregistered} implementations of the Canny edge detector and Hough transform, with the chosen parameters given in \cite{GitLab}.
More elaborate methods may provide better estimates but exceed the scope of this paper. 
The detected lines (dashed) are indicated in Fig.\,\ref{fig:imaging} alongside the resulting mirror sources (dotted) which are computed according to~\cite{Deutschmann23ICC}. 
It is clearly visible that the main portion of power is concentrated at the intersection with the path between mirror sources of the \gls{ula} and the \gls{ura} and thus the radar image does not capture the full extent of the walls. 
\ifdefined\reduceSize  
\else
    However, the physically large extent of our \gls{ula} w.r.t. the propagation distances of interest covers a reasonably large portion of the walls in the resulting radar image in Fig.\,\ref{fig:imaging}.
    This is a feature of the sub-10\,GHz operating frequency range which allows forming physically large apertures.
\fi
\ifthenelse{\equal{\externalizeFigures}{true}}
{
	\tikzexternaldisable    
}
{}
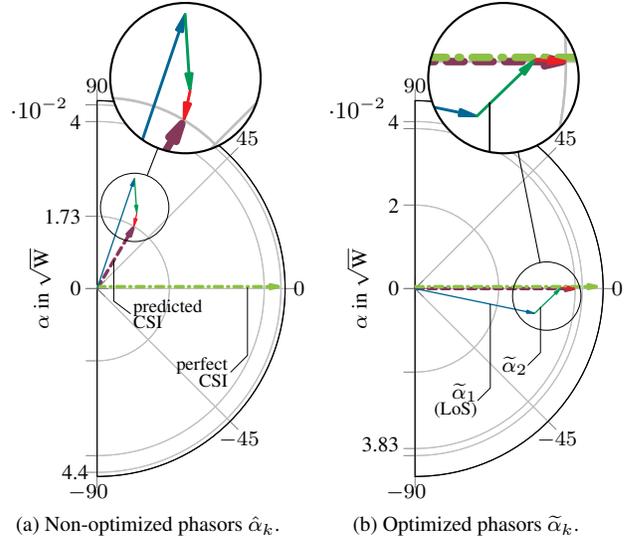
\begin{figure}[t!]
    \setlength{\plotWidth}{0.24\textwidth}
    \begin{subfigure}[b]{0.45\columnwidth}
        \centering
        \definecolor{mycolor1}{rgb}{0.00000,0.44700,0.74100}%
        \definecolor{mycolor2}{rgb}{0.85000,0.32500,0.09800}%
        \definecolor{mycolor3}{rgb}{0.92900,0.69400,0.12500}%
        \definecolor{mycolor4}{rgb}{0.49400,0.18400,0.55600}%
        \definecolor{mycolor5}{rgb}{0.46600,0.67400,0.18800}%
        \definecolor{mycolor6}{rgb}{0.30100,0.74500,0.93300}%
        \definecolor{mycolor7}{rgb}{0.63500,0.07800,0.18400}%
        
        \newcommand{\plotLW}{0.5pt}
        \newcommand{\plotLWh}{1.3pt}

\pgfplotsset{every axis/.append style={
  label style={font=\footnotesize},
  legend style={font=\footnotesize},
  tick label style={font=\footnotesize},
  xticklabel={
    \ifdim \tick pt < 0pt
      \pgfmathparse{abs(\tick)}%
      \llap{$-{}$}\pgfmathprintnumber{\pgfmathresult}
   \else
      \pgfmathprintnumber{\tick}
   \fi
}}}
        
        \begin{tikzpicture}[spy using outlines={circle, magnification=2.2, size=2cm, connect spies}]
        \begin{polaraxis}[
        width=5cm,
        height=5cm,
        scale only axis,
        line cap = round,
        xmin=-90, xmax=90, 
        ymin=0, ymax=0.0450, 
        domain=-90:90,
        ytick={0,0.0173,0.04,0.044},
        yticklabels={,,,4.4},
        yticklabel style={anchor=east},
        every y tick scale label/.style={at={(-0.07,0.02)},color={black!90},text opacity={0}},    
        extra y tick style={grid=none}
        ]
        \end{polaraxis} 
        \begin{polaraxis}[width=5cm,
height=5cm,
scale only axis,
xmin=90, xmax=270, rotate=180, 
        domain=90:91,
        ymin=0, ymax=0.0450,
        y dir=reverse,
        yticklabel style={anchor=east,xshift =-0.1cm},
        every y tick scale label/.style={at={(0.07,0.02)},font={\small},draw opacity={10}},    
        xticklabels={,,},
        ytick={0,0.0173,0.04,0.044},
        yticklabels={0,1.73,4,},
        extra x tick style={grid=none}
        ]
        \end{polaraxis}
        \begin{axis}[%
width=2.5cm,
height=5cm,
scale only axis,
line cap = round,
ylabel={$\alpha$ in $\sqrt{\text{W}}$},
ylabel style={yshift=0.5cm},
x axis line style= { draw opacity=0 },
xticklabels={,,},   
yticklabels={,,},
ticks=none,
every y tick scale label/.style={at={(-0.07,0.02)},color={black!0},draw opacity={0}},    
every x tick scale label/.style={at={(-0.07,0.02)},color={black!0},draw opacity={0}},    
xmin=0, xmax=0.045,
ymin=-0.045, ymax=0.0450
]
\addplot [color=RDmaroon, densely dashed, line width=\plotLWh, forget plot, line cap = round, arrows = {-Stealth[round, inset=0pt, scale=0.85, angle'=20]}] 
  table[row sep=crcr]{%
0	0\\
0.00882107069972402	0.0148408757846523\\
};
\addplot [color=RDlightgreen, dash dot, line width=\plotLWh, forget plot, line cap = round, arrows = {-Stealth[round, inset=0pt, scale=0.85, angle'=20]}] 
  table[row sep=crcr]{%
0	0.000459773900070848\\
0.0437879904829382	0.000459773900070848\\
};
\addplot [color=MidnightBlue, line width=\plotLW, forget plot, line cap = round, arrows = {-Stealth[round, inset=0pt, scale=0.85, angle'=20]}]
  table[row sep=crcr]{%
0	0\\
0.00896477670775902	0.0264040183604106\\
};
\addplot [color=ForestGreen, line width=\plotLW, forget plot, line cap = round, arrows = {-Stealth[round, inset=0pt, scale=0.85, angle'=20]}] 
  table[row sep=crcr]{%
0.00896477670775902	0.0264040183604106\\
0.00950567543028335	0.0180256562135982\\
};
\addplot [color=Red, line width=\plotLW, forget plot, line cap = round, arrows = {-Stealth[round, inset=0pt, scale=0.85, angle'=20]}] 
  table[row sep=crcr]{%
0.00950567543028335	0.0180256562135982\\
0.00882107069972402	0.0148408757846522\\
};

\draw[color=black, line width=0.3pt]    
(0.004,0.0067) -- %
(0.004,-0.003) -- %
 (0.008,-0.006) 
 node[right, xshift=-2pt,text width=10em, text centered,align=left]{\scriptsize predicted \\[-1.5mm] \scriptsize CSI}; 

\draw[color=black, line width=0.3pt]    
(0.036,0.0002) -- %
(0.036,-0.016) -- %
(0.032,-0.020) 
node[left, xshift=2pt,text width=10em, text centered,align=right]{\scriptsize perfect \\[-1.5mm] \scriptsize CSI};

\end{axis}
\spy [black] on (0.50,3.58) in node[fill=white] at (1.175,5.3); 
\end{tikzpicture}
        \captionsetup{justification=centering}
        \vspace{-5mm}
        \caption{Non-optimized phasors $\hat{\alpha}_k$.}
        \label{fig:alpha-non-opt}
    \end{subfigure}
    \hspace{2mm} %
	\begin{subfigure}[b]{0.45\columnwidth}
	\centering
        \definecolor{mycolor1}{rgb}{0.00000,0.44700,0.74100}%
        \definecolor{mycolor2}{rgb}{0.85000,0.32500,0.09800}%
        \definecolor{mycolor3}{rgb}{0.92900,0.69400,0.12500}%
        \definecolor{mycolor4}{rgb}{0.49400,0.18400,0.55600}%
        \definecolor{mycolor5}{rgb}{0.46600,0.67400,0.18800}%
        \definecolor{mycolor6}{rgb}{0.30100,0.74500,0.93300}%
        \definecolor{mycolor7}{rgb}{0.63500,0.07800,0.18400}%
        
        \newcommand{\plotLW}{0.5pt}
        \newcommand{\plotLWh}{1.3pt}

\pgfplotsset{every axis/.append style={
  label style={font=\footnotesize},
  legend style={font=\footnotesize},
  tick label style={font=\footnotesize},
  xticklabel={
    \ifdim \tick pt < 0pt
      \pgfmathparse{abs(\tick)}%
      \llap{$-{}$}\pgfmathprintnumber{\pgfmathresult}
   \else
      \pgfmathprintnumber{\tick}
   \fi
}}}
        
        \begin{tikzpicture}[spy using outlines={circle, magnification=2.2, size=2cm, connect spies}]
        \begin{polaraxis}[
        width=5cm,
        height=5cm,
        scale only axis,
        line cap = round,
        xmin=-90, xmax=90, 
        ymin=0, ymax=0.0450, 
        domain=-90:90,
        ytick={0,0.02,0.0383,0.04},
        yticklabels={,,,3.83,},
        yticklabel style={anchor=east,xshift =-0.1cm,yshift=0.15cm},
        every y tick scale label/.style={at={(-0.07,0.02)},color={black!0},text opacity={0}},    
        extra y tick style={grid=none}
        ]
        \end{polaraxis} 
        \begin{polaraxis}[width=5cm,
        height=5cm,
        scale only axis,
        line cap = round,
        xmin=90, xmax=270, rotate=180, 
        domain=90:91,
        ymin=0, ymax=0.0450,
        y dir=reverse,
        yticklabel style={anchor=east,xshift =-0.1cm},
        every y tick scale label/.style={at={(0.07,0.02)},font={\small},draw opacity={0}},    
        xticklabels={,,},
        ytick={0,0.02,0.0383,0.04},
        yticklabels={0,2,,4},
        extra x tick style={grid=none}
        ]
        \end{polaraxis}
        %
        %
        %
        %
        %
        %
        %
        \begin{axis}[%
width=2.5cm,
height=5cm,
scale only axis,
line cap = round,
ylabel={$\alpha$ in $\sqrt{\text{W}}$},
ylabel style={yshift=0.5cm},
x axis line style= { draw opacity=0 },
xticklabels={,,},   
yticklabels={,,},
ticks=none,
every y tick scale label/.style={at={(-0.07,0.02)},color={black!0},draw opacity={0}},    
every x tick scale label/.style={at={(-0.07,0.02)},color={black!0},draw opacity={0}},    
xmin=0, xmax=0.045,
ymin=-0.045, ymax=0.0450
]
\addplot [color=RDmaroon, densely dashed, line width=\plotLWh, forget plot, line cap = round, arrows = {-Stealth[round, inset=0pt, scale=0.85, angle'=20]}] 
  table[row sep=crcr]{%
0	0\\
0.0383452635526306	0\\
};
\addplot [color=RDlightgreen, dash dot, line width=\plotLWh, forget plot, line cap = round, arrows = {-Stealth[round, inset=0pt, scale=0.85, angle'=20]}] 
  table[row sep=crcr]{%
0	0.000459773900070848\\
0.0437879904829382	0.000459773900070848\\
};
\addplot [color=MidnightBlue, line width=\plotLW, forget plot, line cap = round, arrows = {-Stealth[round, inset=0pt, scale=0.85, angle'=20]}] 
  table[row sep=crcr]{%
0	0\\
0.028611398640816	-0.00593526331580027\\
};
\addplot [color=ForestGreen, line width=\plotLW, forget plot, line cap = round, arrows = {-Stealth[round, inset=0pt, scale=0.85, angle'=20]}] 
  table[row sep=crcr]{%
0.028611398640816	-0.00593526331580028\\
0.0349364092475386	0.000180365319614927\\
};
\addplot [color=Red, line width=\plotLW, forget plot, line cap = round, arrows = {-Stealth[round, inset=0pt, scale=0.85, angle'=20]}] 
  table[row sep=crcr]{%
0.0349364092475386	0.000180365319614927\\
0.0383452635526306	-0\\
};




 \draw[color=black, line width=0.3pt]    
(0.018,-0.0038) -- %
(0.018,-0.0238) -- %
(0.015,-0.0268) 
node[left, xshift=3pt,text width=10em, text centered,align=right]{\footnotesize$\widetilde{\alpha}_1$ \\[-1.5mm] \scriptsize(LoS)};

 \draw[color=black, line width=0.3pt]    
(0.03,-0.0045) -- %
(0.03,-0.01535) -- %
(0.027,-0.01835) 
node[left, xshift=3pt,text width=10em, text centered,align=right]{\footnotesize $\widetilde{\alpha}_2$};

\end{axis}
\spy [black] on (1.75,2.4) in node[fill=white] at (1.175,5.3); 
\end{tikzpicture}
        \captionsetup{justification=centering}
        \vspace{-5mm}
        \caption{Optimized phasors $\widetilde{\alpha}_k$.}
        \label{fig:alpha-opt}
    \end{subfigure}%
    \vspace{-1mm}%
    \caption{Phasors $\alpha_k$ in the complex polar plane computed on the ``true'' (measured) channel vector $\bm{h}$ (dash-dotted) with beamforming weights generated from the predicted \gls{smc} channel vectors $\widetilde{\bm{h}}_k$ (solid) 
    and $P_\mrm{t}=\SI{1}{\watt}$. The phase optimization in Section~\ref{sec:phase-optimization} aligns the \gls{smc} beam phases and maximizes the sum-phasor (dashed) received by the \gls{ue}.
    }\label{fig:polar-plots}
    \vspace{-6mm}
\end{figure}
\ifthenelse{\equal{\externalizeFigures}{true}}
{
	\tikzexternalenable     
}
{}

\section{Wireless Power Transfer}\label{sec:wpt}%
Using the results obtained in Section~\ref{sec:sensing}, we show how the inferred geometric environment information can be exploited for the exemplary application of narrow-band \gls{wpt}, a promising service to be provided by future radio infrastructures.

\subsection{Geometry-based Beamforming}\label{sec:geometry-based-bf}%
In the following, we assume that, instead of a \gls{ura}, a single-antenna \gls{ue} is placed at the location $\bm{p}_\textsc{ue}$, selected as the center of gravity of the \gls{ura}.
This allows using the collected measurement data in a \gls{miso} configuration. 
We aim to transmit power to the \gls{ue} based solely on the assumed known location of the \gls{ue} and the inferred locations of the mirror sources and the \gls{los}, i.e., using $K=3$ \glspl{smc}. 
We use a geometry-based beamformer at the chosen frequency of $f_\mrm{c} = \SI{3.79}{\giga\hertz}$ and compute beamforming weights using \gls{mrt} as
\begin{align}\label{eq:sum-weights-non-opt}
    \bm{w}  
            = \frac{
                \sum_{k=1}^K {\bm{w}}_k
            }{\lVert 
                \sum_{k=1}^K {\bm{w}}_k
            \rVert}
    \hspace{0.4cm} \text{with} \hspace{0.4cm} 
    {\bm{w}}_k=\frac{\widetilde{\bm{h}}_k^*(\bm{p}_\textsc{ue},f_\mrm{c})}{\lVert \widetilde{\bm{h}}(\bm{p}_\textsc{ue},f_\mrm{c}) \rVert}
\end{align}
where the predicted channel vector $\widetilde{\bm{h}}(\bm{p}_\textsc{ue},f_\mrm{c})$ is the superposition of the $K=3$ predicted \gls{smc} channel vectors $\widetilde{\bm{h}}_k(\bm{p}_\textsc{ue},f_\mrm{c})$, computed using~\eqref{eq:smc-channel-vector} and the estimated mirror source locations.
We can 
compute 
the phasors for each \gls{smc} $k$ using \eqref{eq:y_rx} as
\begin{align}\label{eq:y_rx_approx}
    \hat{\alpha}_k = \bm{h}^\trp \bm{w}_k \sqrt{P_\mrm{t}} 
\end{align}
with the assumed ``true'' (measured) channel vector $\bm{h}$ to quantify the contribution of each mirror source on the sum-phasor $\hat{\alpha}=\sum_{k=1}^K \hat{\alpha}_k$ received by the \gls{ue}. 
Note that the channel vectors $\bm{h}_k$ are not independent and thus the computed phasors $\hat{\alpha}_k$ only approximate the amplitudes of the $k$\textsuperscript{th} \gls{smc}.
Fig.\,\ref{fig:alpha-non-opt} shows that the \gls{smc} phasors $\hat{\alpha}_2$ and $\hat{\alpha}_3$ are not well aligned with the \gls{los} phasor $\hat{\alpha}_1$ as a result of uncertainty in the estimated mirror source locations. 
The \gls{smc} beams interfere destructively at the \gls{ue}, such that the path gain is only $PG \approx \SI{-35.3}{\dB}$ using our predicted weights in \eqref{eq:sum-weights-non-opt}.

\subsection{Optimization of Beam Phases}\label{sec:phase-optimization}%
\setlength{\figurewidth}{0.425\linewidth}
\setlength{\hspaceing}{-1.56cm}
\begin{figure}[tb]
     \hspace{-3mm}
     \begin{subfigure}[t]{0.48\linewidth}
        \vskip 0pt	
            \includegraphics[width=\linewidth]{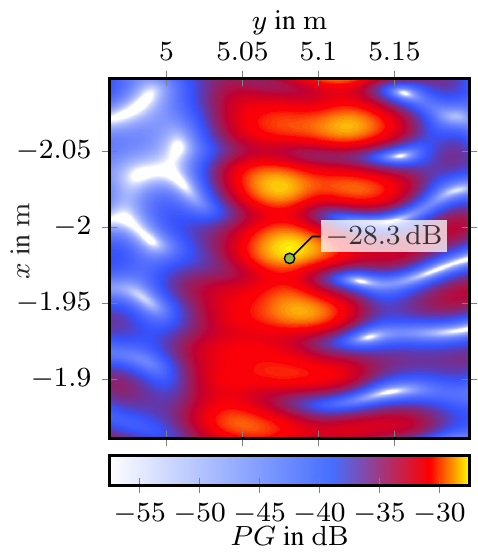}
            \vspace{-6mm}
         \caption{Using all sources $k\in\{1,2,3\}$.}
         \label{fig:beamforming-k123}
     \end{subfigure}%
     \hspace{0.5cm}%
     \begin{subfigure}[t]{0.48\linewidth}
        \vskip 0pt	
         \centering
            \includegraphics[width=0.865\linewidth]{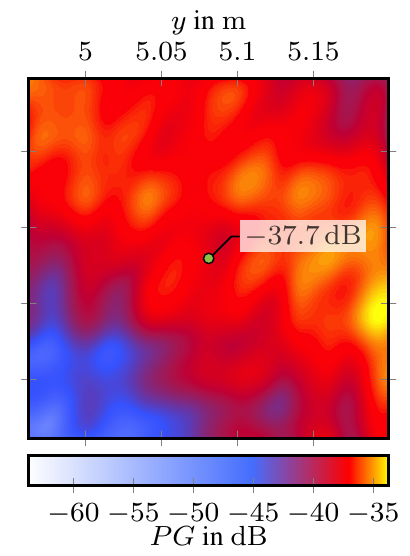}
            \vspace{-2.5mm}
            \caption{Using source $k=3$ only.}
            \label{fig:beamforming-k3}
     \end{subfigure}%
     \vspace{-1mm}%
     \caption{Measured $PG$ distribution across the synthetic \gls{ura} (interpolated) when applying geometry-based beamforming:
     Channel vectors $\bm{h}_k$ are predicted using the estimated geometric environment information from Fig.\,\ref{fig:imaging}.}%
     \label{fig:beamforming}%
     \vspace{-5mm}
\end{figure}
To compensate for geometric uncertainties in the environment model, we employ an optimization of \gls{smc} beam phases that we proposed in~\cite[eq.\,(14)]{Deutschmann23ICC}. 
The objective is to find optimal phase shifts $\widetilde{\varphi}_k$ applied to the weights $\bm{w}_k$ 
such that the path gain at the \gls{ue} is maximized. 
Note that the number of beam phases to be optimized is $K-1$, i.e., the phase of one beam (e.g., the \gls{los} beam) can be kept constant and all other beam phases are optimized. 
Fig.\,\ref{fig:alpha-opt} shows the corresponding optimized phasors $\tilde{\alpha}_k$. 
After the optimization, the \gls{ue} receives a sum-phasor that translates to a path gain $PG \approx \SI{-28.3}{\dB}$ using our predicted \gls{csi}, which gets reasonably close to the maximum path gain $PG_\textsc{max} \approx \SI{-27.2}{\dB}$ achievable with perfect \gls{csi}.
Fig.\,\ref{fig:beamforming-k123} shows the $PG$ distribution across the aperture of the \gls{ura} given the optimized beamforming weights $\widetilde{\bm{w}}$.
\ifdefined\reduceSize  
\else
    A strong standing wave pattern is visible in Fig.\,\ref{fig:beamforming-k123}, originating from the wall ``behind'' the \gls{ue}, i.e., (mirror) sources $k=2$ and $k=1$ are located on opposite sides of the \gls{ue}.
    This is a particular problem of performing \gls{wpt} in indoor environments as we have demonstrated in~\cite{Deutschmann22ICC}.
\fi
Fig.\,\ref{fig:beamforming-k3} shows the resulting $PG$ distribution when using the mirror source $k=3$ only, illustrating how the geometric model uncertainty impacts the location of the resulting \gls{smc} beam. 
\ifdefined\reduceSize  
     At the same time, it confirms that the reflection from the corresponding metal wall is reasonably specular as our \gls{smc} channel model results in a clearly visible beam originating from the location of the third mirror source.
\else
    The right metal wall has been detected too close to the actual physical \gls{ula}, and thus the image source $k=3$ is located too far left.
    The resulting geometrically constructed beam encloses a too narrow angle with the metal wall and is therefore not well aligned with the \gls{ue}.
    Fig.\,\ref{fig:beamforming-k3} further confirms that the reflection from the corresponding metal wall is reasonably specular as our \gls{smc} channel model results in a clearly visible beam originating from the location of the third mirror source.
\fi
\ifdefined\reduceSize  
    ~ 
\else
    \section{Dual-band Operation}
    \begin{figure}
      \centering
      \includegraphics[width=\linewidth]{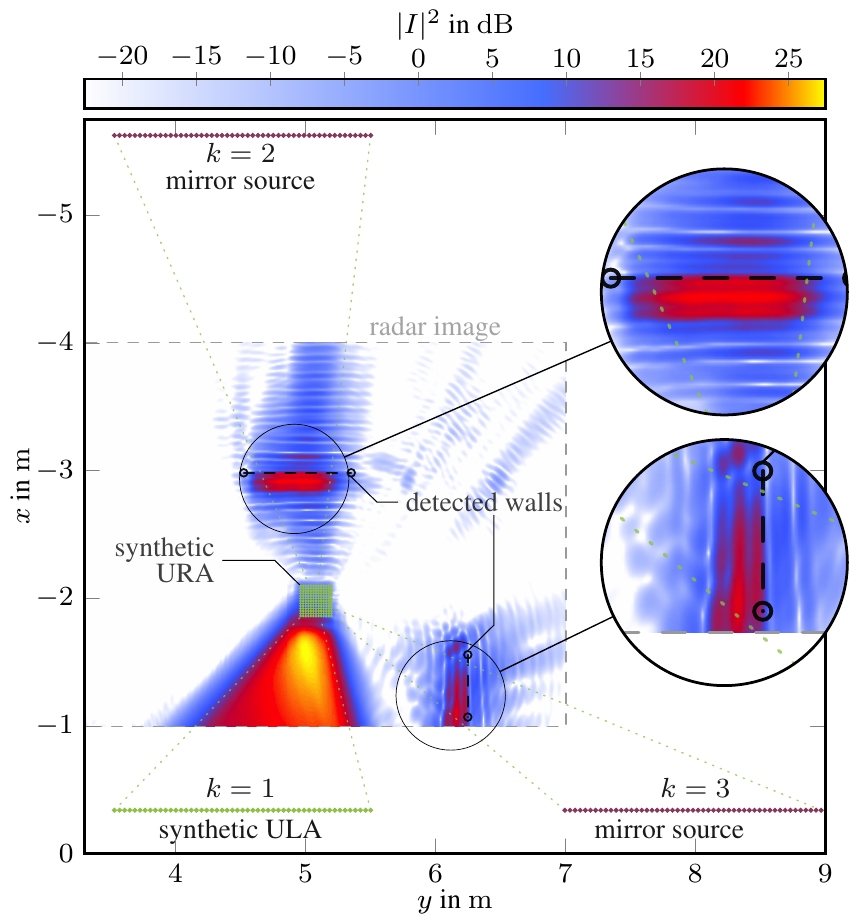}
      \vspace{-7mm}
      \caption{%
      The obtained bistatic \gls{mimo} radar image generated with the dual-band operation. 
      The walls are inferred at ``ripples'' in the radar image at some distance from the locations that would be estimated using the full bandwidth (see Fig.~\ref{fig:imaging}).
      }
        \label{fig:DB-imaging}
        \vspace{-6mm}
    \end{figure} %
 
    Future distributed radio infrastructures like RadioWeaves may not have a large frequency band of 
    ${3}$-$\SI{10}{\giga\hertz}$
    available. 
    However, a dual-band operation may be a suitable alternative to provide sufficient imaging results for inferring walls.
    To test the performance of a dual-band operation, 
    we restrict the measured bandwidth to a \SI{100}{\mega\hertz} band centered around \SI{3.79}{\giga\hertz} and a \SI{1.2}{\giga\hertz} band centered around \SI{6.5}{\giga\hertz} (a frequency band designated for Wi-Fi\,6E in the U.S., South Korea, Brazil, and Canada~\cite{Mehrnoush2022WiFi6E}) and repeat the imaging procedure.
    The resulting radar image is shown in Fig.\,\ref{fig:DB-imaging}.
    As can be observed from the figure, walls are inferred at ``ripples'' in the radar image at some distance from the previously estimated locations given the full bandwidth.
    However, Fig.\,\ref{fig:DB-polar-plots} shows that the phase-optimization method still attains a reasonable efficiency of $PG\approx\SI{-28.2}{\dB}$ for wireless power transfer. 
    More sophisticated schemes may be better able to estimate wall locations and provide better results.
    \ifthenelse{\equal{\externalizeFigures}{true}}
{
    \tikzexternaldisable    
}
{}
    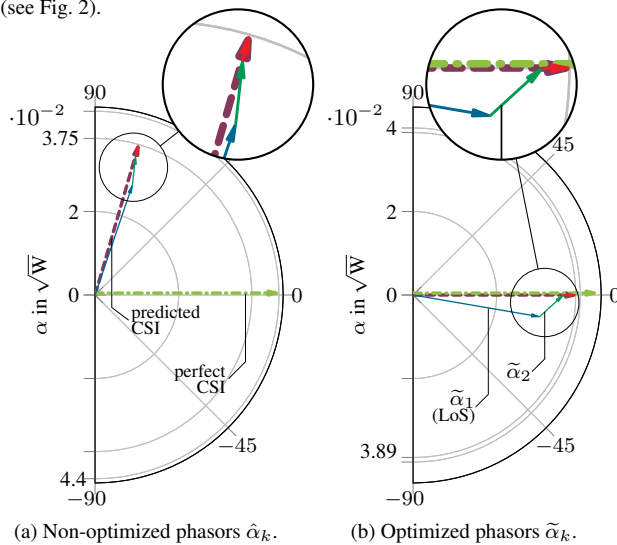
\begin{figure}[t!]
        \setlength{\plotWidth}{0.24\textwidth}
        \begin{subfigure}[b]{0.45\columnwidth}
            \centering
            \definecolor{mycolor1}{rgb}{0.00000,0.44700,0.74100}%
        \definecolor{mycolor2}{rgb}{0.85000,0.32500,0.09800}%
        \definecolor{mycolor3}{rgb}{0.92900,0.69400,0.12500}%
        \definecolor{mycolor4}{rgb}{0.49400,0.18400,0.55600}%
        \definecolor{mycolor5}{rgb}{0.46600,0.67400,0.18800}%
        \definecolor{mycolor6}{rgb}{0.30100,0.74500,0.93300}%
        \definecolor{mycolor7}{rgb}{0.63500,0.07800,0.18400}%
        
        \newcommand{\plotLW}{0.5pt}
        \newcommand{\plotLWh}{1.3pt}

\pgfplotsset{every axis/.append style={
  label style={font=\footnotesize},
  legend style={font=\footnotesize},
  tick label style={font=\footnotesize},
  xticklabel={
    \ifdim \tick pt < 0pt
      \pgfmathparse{abs(\tick)}%
      \llap{$-{}$}\pgfmathprintnumber{\pgfmathresult}
   \else
      \pgfmathprintnumber{\tick}
   \fi
}}}
        
        \begin{tikzpicture}[spy using outlines={circle, magnification=2.2, size=2cm, connect spies}]
        \begin{polaraxis}[
        width=5cm,
        height=5cm,
        scale only axis,
        line cap = round,
        xmin=-90, xmax=90, 
        ymin=0, ymax=0.0450, 
        domain=-90:90,
        ytick={0,0.02,0.0375,0.044},
        yticklabels={,,,4.4},
        yticklabel style={anchor=east},
        every y tick scale label/.style={at={(-0.07,0.02)},color={black!90},text opacity={0}},    
        extra y tick style={grid=none}
        ]
        \end{polaraxis} 
        \begin{polaraxis}[width=5cm,
height=5cm,
scale only axis,
xmin=90, xmax=270, rotate=180, 
        domain=90:91,
        ymin=0, ymax=0.0450,
        y dir=reverse,
        yticklabel style={anchor=east,xshift =-0.1cm},
        every y tick scale label/.style={at={(0.07,0.02)},font={\small},draw opacity={10}},    
        xticklabels={,,},
        ytick={0,0.02,0.0375,0.044},
        yticklabels={0,2,3.75,},
        extra x tick style={grid=none}
        ]
        \end{polaraxis}
        \begin{axis}[%
width=2.5cm,
height=5cm,
scale only axis,
line cap = round,
ylabel={$\alpha$ in $\sqrt{\text{W}}$},
ylabel style={yshift=0.5cm},
x axis line style= { draw opacity=0 },
xticklabels={,,},   
yticklabels={,,},
ticks=none,
every y tick scale label/.style={at={(-0.07,0.02)},color={black!0},draw opacity={0}},    
every x tick scale label/.style={at={(-0.07,0.02)},color={black!0},draw opacity={0}},    
xmin=0, xmax=0.045,
ymin=-0.045, ymax=0.0450
]
\addplot [color=RDmaroon, densely dashed, line width=\plotLWh, forget plot, line cap = round, arrows = {-Stealth[round, inset=0pt, scale=0.85, angle'=20]}] 
  table[row sep=crcr]{%
0	0\\
0.010397759924368	0.0358505439565327\\
};
\addplot [color=RDlightgreen, dash dot, line width=\plotLWh, forget plot, line cap = round, arrows = {-Stealth[round, inset=0pt, scale=0.85, angle'=20]}] 
  table[row sep=crcr]{%
0	0.000459773900070848\\
0.0437879904829382	0.000459773900070848\\
};
\addplot [color=MidnightBlue, line width=\plotLW, forget plot, line cap = round, arrows = {-Stealth[round, inset=0pt, scale=0.85, angle'=20]}]
  table[row sep=crcr]{%
0	0\\
0.00894006175324525	0.0263312252352686\\
};
\addplot [color=ForestGreen, line width=\plotLW, forget plot, line cap = round, arrows = {-Stealth[round, inset=0pt, scale=0.85, angle'=20]}] 
  table[row sep=crcr]{%
0.00894006175324526	0.0263312252352686\\
0.00970345139767841	0.0332678901594341\\
};
\addplot [color=Red, line width=\plotLW, forget plot, line cap = round, arrows = {-Stealth[round, inset=0pt, scale=0.85, angle'=20]}] 
  table[row sep=crcr]{%
0.00970345139767841	0.0332678901594341\\
0.010397759924368	0.0358505439565327\\
};

\draw[color=black, line width=0.3pt]    
(0.004,0.0135) -- %
(0.004,-0.003) -- %
 (0.008,-0.006) 
 node[right, xshift=-2pt,text width=10em, text centered,align=left]{\scriptsize predicted \\[-1.5mm] \scriptsize CSI}; 

\draw[color=black, line width=0.3pt]    
(0.036,0.0002) -- %
(0.036,-0.016) -- %
(0.032,-0.020) 
node[left, xshift=2pt,text width=10em, text centered,align=right]{\scriptsize perfect \\[-1.5mm] \scriptsize CSI};

\end{axis}
\spy [black] on (0.51,4.2) in node[fill=white] at (1.9,5.3); 
\end{tikzpicture}
            \captionsetup{justification=centering}
            \vspace{-5mm}
            \caption{Non-optimized phasors $\hat{\alpha}_k$.}
            \label{fig:DB-alpha-non-opt}
        \end{subfigure}
        \hspace{2mm} %
    	\begin{subfigure}[b]{0.45\columnwidth}
    	\centering
            \definecolor{mycolor1}{rgb}{0.00000,0.44700,0.74100}%
        \definecolor{mycolor2}{rgb}{0.85000,0.32500,0.09800}%
        \definecolor{mycolor3}{rgb}{0.92900,0.69400,0.12500}%
        \definecolor{mycolor4}{rgb}{0.49400,0.18400,0.55600}%
        \definecolor{mycolor5}{rgb}{0.46600,0.67400,0.18800}%
        \definecolor{mycolor6}{rgb}{0.30100,0.74500,0.93300}%
        \definecolor{mycolor7}{rgb}{0.63500,0.07800,0.18400}%
        
        \newcommand{\plotLW}{0.5pt}
        \newcommand{\plotLWh}{1.3pt}

\pgfplotsset{every axis/.append style={
  label style={font=\footnotesize},
  legend style={font=\footnotesize},
  tick label style={font=\footnotesize},
  xticklabel={
    \ifdim \tick pt < 0pt
      \pgfmathparse{abs(\tick)}%
      \llap{$-{}$}\pgfmathprintnumber{\pgfmathresult}
   \else
      \pgfmathprintnumber{\tick}
   \fi
}}}
        
        \begin{tikzpicture}[spy using outlines={circle, magnification=2.2, size=2cm, connect spies}]
        \begin{polaraxis}[
        width=5cm,
        height=5cm,
        scale only axis,
        line cap = round,
        xmin=-90, xmax=90, 
        ymin=0, ymax=0.0450, 
        domain=-90:90,
        ytick={0,0.02,0.0389,0.04},
        yticklabels={,,,3.89,},
        yticklabel style={anchor=east,xshift =-0.1cm,yshift=0.15cm},
        every y tick scale label/.style={at={(-0.07,0.02)},color={black!0},text opacity={0}},    
        extra y tick style={grid=none}
        ]
        \end{polaraxis} 
        \begin{polaraxis}[width=5cm,
        height=5cm,
        scale only axis,
        line cap = round,
        xmin=90, xmax=270, rotate=180, 
        domain=90:91,
        ymin=0, ymax=0.0450,
        y dir=reverse,
        yticklabel style={anchor=east,xshift =-0.1cm},
        every y tick scale label/.style={at={(0.07,0.02)},font={\small},draw opacity={0}},    
        xticklabels={,,},
        ytick={0,0.02,0.0389,0.04},
        yticklabels={0,2,,4},
        extra x tick style={grid=none}
        ]
        \end{polaraxis}
        %
        %
        %
        %
        %
        %
        %
        \begin{axis}[%
width=2.5cm,
height=5cm,
scale only axis,
line cap = round,
ylabel={$\alpha$ in $\sqrt{\text{W}}$},
ylabel style={yshift=0.5cm},
x axis line style= { draw opacity=0 },
xticklabels={,,},   
yticklabels={,,},
ticks=none,
every y tick scale label/.style={at={(-0.07,0.02)},color={black!0},draw opacity={0}},    
every x tick scale label/.style={at={(-0.07,0.02)},color={black!0},draw opacity={0}},    
xmin=0, xmax=0.045,
ymin=-0.045, ymax=0.0450
]
\addplot [color=RDmaroon, densely dashed, line width=\plotLWh, forget plot, line cap = round, arrows = {-Stealth[round, inset=0pt, scale=0.85, angle'=20]}] 
  table[row sep=crcr]{%
0	0\\
0.0389170923593081	0\\
};
\addplot [color=RDlightgreen, dash dot, line width=\plotLWh, forget plot, line cap = round, arrows = {-Stealth[round, inset=0pt, scale=0.85, angle'=20]}] 
  table[row sep=crcr]{%
0	0.000459773900070848\\
0.0437879904829382	0.000459773900070848\\
};
\addplot [color=MidnightBlue, line width=\plotLW, forget plot, line cap = round, arrows = {-Stealth[round, inset=0pt, scale=0.85, angle'=20]}] 
  table[row sep=crcr]{%
0	0\\
0.030271026999224	-0.00519887840617005\\
};
\addplot [color=ForestGreen, line width=\plotLW, forget plot, line cap = round, arrows = {-Stealth[round, inset=0pt, scale=0.85, angle'=20]}] 
  table[row sep=crcr]{%
0.030271026999224	-0.00519887840617005\\
0.0359631911096091	-1.50927969711639e-06\\
};
\addplot [color=Red, line width=\plotLW, forget plot, line cap = round, arrows = {-Stealth[round, inset=0pt, scale=0.85, angle'=20]}] 
  table[row sep=crcr]{%
0.0359631911096091	-1.50927969711639e-06\\
0.0389170923593081	6.93889390390723e-18\\
};




 \draw[color=black, line width=0.3pt]    
(0.018,-0.0033) -- %
(0.018,-0.0238) -- %
(0.015,-0.0268) 
node[left, xshift=3pt,text width=10em, text centered,align=right]{\footnotesize$\widetilde{\alpha}_1$ \\[-1.5mm] \scriptsize(LoS)};

 \draw[color=black, line width=0.3pt]    
(0.0315,-0.0041) -- %
(0.0315,-0.01531) -- %
(0.0285,-0.01831) 
node[left, xshift=3pt,text width=10em, text centered,align=right]{\footnotesize $\widetilde{\alpha}_2$};

\end{axis}
\spy [black] on (1.75,2.4) in node[fill=white] at (1.175,5.3); 
\end{tikzpicture}
            \captionsetup{justification=centering}
            \vspace{-5mm}
            \caption{Optimized phasors $\widetilde{\alpha}_k$.}
            \label{fig:DB-alpha-opt}
        \end{subfigure}%
        \vspace{-1mm}%
        \caption{Phasors $\alpha_k$ in the complex polar plane generated with the dual-band operation. 
        A reasonable efficiency of $PG\approx\SI{-28.2}{\dB}$ is attained
        after the phase optimization in Section~\ref{sec:phase-optimization}, despite deviations in the estimated wall locations.
        }\label{fig:DB-polar-plots}
        \vspace{-6mm}
    \end{figure}

        \ifthenelse{\equal{\externalizeFigures}{true}}
{
    \tikzexternalenable     
}
{}
\fi
\section{Conclusion}\label{sec:conclusion}%
We evaluated a bistatic \gls{mimo} radar imaging scheme on \gls{sa} measurements at sub-10\,GHz frequencies with the aim of detecting specularly reflecting surfaces in the environment. 
%
In contrast to imaging, single snapshot-based estimation of mirror sources 
cannot resolve whether multipath components originate from specularly reflecting surfaces (i.e., SMCs) or point scatterers. 
For some applications, such as WPT, the former can be particularly exploited. 
The imaging-based method would detect specular surfaces rather than point scatterers.
%
We were able to detect sharp edges at the locations of the surfaces and not their corresponding virtual mirror sources, which is generally a characteristic of diffuse 
reflections. 
We extracted walls as geometric environment features 
and constructed channel vectors for beamforming solely based on geometric information, thus performing sensing-aided beam prediction for \gls{wpt}.
\ifdefined\reduceSize  
\else
    To achieve reasonable imaging results, we have found that a large bandwidth is needed given the simple imaging scheme used in this work. 
    The dual-band operation performs worse in inferring the walls (see Fig.\,\ref{fig:DB-imaging}) but still showed a reasonable performance for the given application of \gls{wpt}.
    Thus it may be a suitable approach for future distributed radio infrastructures, especially when being used with more elaborate estimation schemes. 
\fi
After optimizing the \gls{smc} beam phases, our geometry-based beamformer suffers a loss of only \SI{1.1}{\dB} when compared with perfect \gls{csi}.

\bibliographystyle{IEEEbib}
\bibliography{IEEEabrv,ICASSP_2023}

\end{document}